\documentclass[aps,prd,twocolumn,nofootinbib,showpacs]{revtex4-1}

\usepackage{graphicx}

\usepackage[colorlinks=true,linkcolor=blue,citecolor=blue, urlcolor=blue]{hyperref}

\begin{document}

\title{The Higgs-Boson Decay $H\to gg$ up to $\alpha_s^5$-Order \\ under the Minimal Momentum Space Subtraction Scheme}

\author{Dai-Min Zeng$^{1,2}$}

\author{Sheng-Quan Wang$^{1,2,3}$}

\author{Xing-Gang Wu$^{1,2}$}
\email{wuxg@cqu.edu.cn}

\author{Jian-Ming Shen$^{1,2}$}

\address{$^1$Department of Physics, Chongqing University, Chongqing 401331, P.R. China \\ $^2$Institute of Theoretical Physics, Chongqing University, Chongqing 401331, P.R. China \\
$^3$School of Science, Guizhou Minzu University, Guiyang 550025, P.R. China}

\begin{abstract}

We make a detailed study on the Higgs-boson decay width $\Gamma(H\to gg)$ up to $\alpha_s^5$-order under the minimal momentum space subtraction (mMOM) scheme. A major uncertainty of a finite-order perturbative QCD prediction is the perceived ambiguity in setting the renormalization scale. In the present paper, to achieve a precise pQCD prediction without renormalization scale uncertainty, we adopt the Principle of Maximum Conformality (PMC) to set the renormalization scale of the process. The PMC has solid theoretical foundation, which is based on the renormalization group invariance and utilizes the renormalization group equation to fix the renormalization scale of the process. The key point of the PMC applications is how to correctly set the $\{\beta_i\}$-terms of the process such that to achieve the correct $\alpha_s$-running behavior at each perturbative order. It is found that the ambiguities in dealing with the $\{\beta_i\}$-terms of the decay width $\Gamma(H\to gg)$ under the $\overline{\rm MS}$-scheme can be avoided by using the physical mMOM-scheme. For the purpose, as the first time, we provide the PMC scale-setting formulas within the mMOM-scheme up to four-loop level. By using PMC, it is found that a more reliable pQCD prediction on $\Gamma(H\to gg)$ can indeed be achieved under the mMOM-scheme. As a byproduct, the convergence of the resultant pQCD series has also been greatly improved due to the elimination of renormalon terms. By taking the newly measured Higgs mass, $M_H=125.09\pm 0.21\pm 0.11$ GeV, our PMC prediction of the decay width is, $\Gamma(H\to gg)|_{\rm mMOM, PMC} = 339.3 \pm 1.7^{+3.7}_{-2.4}$ keV, in which the first error is from the Higgs mass uncertainty and the second error is the residual renormalization scale dependence by varying the initial renormalization scale $\mu_r\in[M_H/2,4M_H]$.


\end{abstract}

\maketitle

\section{Introduction}

Both the ATLAS and the CMS collaborations at the Large Hadron Collider (LHC) have discovered a new gauge boson whose properties are compatible with the Standard Model (SM) Higgs boson~\cite{higgs1, higgs2}. This discovery has initiated a new era of precision studies of the Higgs-boson properties. Among the Higgs-boson decay channels, the decay $H\to gg$ plays an important role in Higgs phenomenology. The next-to-leading order (NLO)~\cite{hgg1, hgg2, hgg3, hgg4, hgg5, hgg6}, the next-to-next-leading order (N$^2$LO)~\cite{K.G.Chetyrkin1, K.G.Chetyrkin2}, and the next-to-next-to-next-leading order (N$^3$LO)~\cite{P.A.Baikov} QCD corrections to the total decay width $\Gamma(H\to gg)$ under the modified minimal subtraction scheme ($\mathrm{\overline{MS}}$-scheme) have been done in the literature. Those improvements provide us great chances for achieving precise perturbative QCD (pQCD) predictions on $\Gamma(H\to gg)$.

A key problem in making precise pQCD prediction is how to set the renormalization scale of the running coupling. Conventionally, the total decay width $\Gamma(H\to gg)$ is predicted by choosing the Higgs mass $M_H$ as the renormalization scale and by varying it within a certain range, such as $[M_{H}/2, 2M_{H}]$, to ascertain its uncertainty. However, at any finite order, such a simple choice of scale leads to the well-known renormalization scheme-and-scale ambiguities. There are also uncanceled large logarithms as well as the ``renormalon" terms in the high-orders that diverge as ($n!\beta_i^{n}\alpha_s^n$), which can give sizable contributions to the theoretical estimation and dilute the pQCD convergence. Schematically, one can express the total decay width $\Gamma(H \to gg)$ as $K\times \Gamma_{\rm Born}(H \to gg)$, where $\Gamma_{\rm Born}$ stands for the tree-level or the leading-order (LO) decay width. It has been shown that the QCD correction factor $K$ under the $\overline{\rm MS}$-scheme follows the trends~\cite{P.A.Baikov}
\begin{equation}
K_{\overline{\rm MS}} \sim 1 + 0.65 +0.20 +0.02 +\cdots , \label{msbarK}
\end{equation}
which shows that the NLO and the N$^2$LO corrections are about $65\%$ and $20\%$ of the Born term, indicating a slow pQCD convergence under the $\overline{\rm MS}$-scheme. Moreover, as will be shown later, there are large scale uncertainties for each loop terms under the conventional scale-setting. It is thus helpful to find a proper scale-setting approach to achieve a more reliable pQCD prediction.

A guiding principle for resolving the renormalization scheme-and-scale ambiguities is that physical results must be independent of theoretical conventions. As has been argued in Refs.\cite{PMC1, PMC2, PMC3, PMC4, PMC5}, if one fixes the renormalization scale of pQCD series using Principle of Maximum Conformality (PMC), all non-conformal $\{\beta_i\}$-terms in pQCD series are then resummed into the running coupling, and one thus obtains a unique, scale-fixed, and scheme-independent prediction at any finite order. The PMC has a solid theoretical foundation, satisfying the renormalization group invariance~\cite{Wu:2014iba} and all other self-consistency conditions from renormalization group equation (RGE)~\cite{Brodsky:2012ms}. The PMC provides the general procedure underlies the commensurate scale relations~\cite{CSR}, which ensures the scheme independence of the predictions under various schemes. A demonstration of the scheme independence at any fixed-order has been given with the help of $R_\delta$-scheme~\cite{PMC4, PMC5}, a systematic generalization of the minimal subtraction renormalization scheme. The PMC reduces in Abelian limit to the standard Gell-Mann-Low scale-setting method used in QED~\cite{GellMann:1954fq}. The PMC has been successfully applied to many high-energy processes, cf. a recent review on the importance of proper renormalization scale-setting for QCD testing at high-energy colliders~\cite{Front}, which can also be applied to processes with multiple physical scales~\cite{Wang:2014aqa, Shen:2015cta}. In the paper, we shall adopt the PMC with the goal of eliminating the renormalization scale ambiguity and achieving a more accurate pQCD prediction.

A PMC analysis of the decay $H\to gg$ under the $\mathrm{\overline{MS}}$-scheme has been done in Ref.\cite{Sheng-Quan Wang}. The pQCD convergence of this process cannot be greatly improved as the usual PMC applications, especially its $\mathrm{N^2}$LO contribution is still about $-20\%$ of the total decay width, indicating there is somewhat larger residual scale dependence. However, it is noted that the simple PMC analysis done in Ref.\cite{Sheng-Quan Wang} could be inaccurate. This is because that the present process involves the three-gluon vertex at the lowest order, thus the scale-setting problem should be much more involved~\cite{three_g}. Only those $\{\beta_i\}$-terms that are pertained to the renormalization of running coupling should be absorbed into the running coupling so as to achieve the optimal scales at each perturbative order. Thus, special treatment, though difficult, should be paid for distributing the $\{\beta_i\}$-terms of the process, which is lacking in Ref.\cite{Sheng-Quan Wang}.

To avoid such ambiguity of applying the PMC, we shall first transform the results from the $\overline{\rm MS}$-scheme to the momentum space subtraction scheme (MOM-scheme)~\cite{W.Celmaster3, W.Celmaster4, W.Celmaster1, W.Celmaster2}, which is based on renormalization of the triple-gluon vertex at some symmetric off-shell momentum, and then apply the PMC scale-setting. Unlike the $\mathrm{\overline{MS}}$-scheme, the physical MOM-scheme carries information of the vertex at specific momentum configuration. This external momentum configuration is non-exceptional and there are no infrared issues, thus avoiding the confusion of distinguishing $\{\beta_i\}$-terms. In the literature, as an alternation of the MOM-scheme, the minimal momentum space subtraction scheme (mMOM-scheme) has also been suggested~\cite{L.von Smekal}, which is an extension of the MOM-scheme on the ghost-gluon vertex and allows the strong running coupling to be fixed solely through a determination of the gluon and ghost propagators. The mMOM-scheme can be related to the $\overline{\rm MS}$-scheme at four-loop level, thus the four-loop mMOM $\beta$-function can be determined with the help of the relations between mMOM and $\overline{\rm MS}$-couplings~\cite{L.von Smekal, J.A.Gracey2}. This four-loop mMOM $\beta$-function can then be adopted to get the four-loop mMOM-scheme $\alpha_s$-running behavior and well suit the needs for our present four-loop analysis.

Using the PMC under the MOM-scheme has already been suggested by Refs.\cite{Pomeron1, Pomeron2, Pomeron3, Pomeron4} to deal with QCD BFKL Pomeron. Those papers show that after applying the PMC, the QCD perturbative convergence can be greatly improved and the BFKL Pomeron intercept has a weak dependence on the virtuality of the reggeized gluon, resulting a much better agreement with the experimental data in comparison to the $\overline{\rm MS}$-prediction. In those references, the NLO transition of the $\overline{\rm MS}$-scheme to the MOM-scheme has been explained. At present, we shall provide a higher-order treatment of such scheme transition and present the PMC formulas under the mMOM-scheme up to four-loop level.

The remaining parts of the paper are organized as follows. In Sec.~\ref{sect2}, we present the technology for calculating the total decay width $\Gamma(H\to gg)$ under the mMOM-scheme up to four-loop level. The behavior of the running coupling under mMOM-scheme is presented. Numerical results and discussions are presented in Sec.~\ref{sect3}. A brief summary will be given in Sec.~\ref{sect4}. For convenience, we put the PMC coefficients under the mMOM-scheme and the Landau gauge in the Appendix.

\section{Total Decay Width $\Gamma(H\to gg)$ under the minimal MOM-Scheme} \label{sect2}

In the literature, the pQCD calculation is usually done under the $\mathrm{\overline{MS}}$-scheme. To apply the PMC, we shall transform the pQCD expressions for the decay $H\to gg$ under the $\mathrm{\overline{MS}}$-scheme into those of the mMOM-scheme by using the relation of running coupling between the $\mathrm{\overline{MS}}$-scheme and the mMOM-scheme.

The MOM-scheme is gauge dependent, and three gauges as Landau gauge ($\xi=0$), Feynman gauge ($\xi=1$) and Fried-Yennie gauge ($\xi=3$) are adopted in the literature, where $\xi$ stands for the gauge parameter. The question is much more involved when the gauge parameter $\xi\neq 0$, especially the following suggested extended renormalization group method can not be directly applied and some alterations must be done to obtain a smooth scheme transformation among different running couplings~\footnote{A detailed discussion on the gauge dependence of the high-order mMOM-scheme prediction is in progress.}. For definiteness, we shall adopt the Landau gauge ($\xi=0$) to do our calculation.

\subsection{The running coupling and the $\beta$-function to four-loop level under the mMOM-scheme }

The scale-running behavior of the running coupling is controlled by the renormalization group equation or the $\beta$-function
\begin{equation} \label{RGE_A}
\mu_r^2\frac{da_\mathrm{A}(\mu_r^2)}{d\mu_r^2} =\beta^\mathrm{A}(a_\mathrm{A})=-\sum_{i=0}^\infty \beta^{A}_ia_\mathrm{A}^{i+2},
\end{equation}
where the symbol ``A" stands for an arbitrary renormalization scheme, $a_{\mathrm{A}}=\alpha_{s,\mathrm{A}}/4\pi$ with $\alpha_{s,\mathrm{A}}$ being the strong running coupling under the $\mathrm{A}$-scheme.

By taking the same integral constant as that of Refs.\cite{Bardeen:1978yd, Furmanski:1981cw}, the solution of Eq.(\ref{RGE_A}) over the power series of $1/L_A$ can be written as,
\begin{widetext}
\begin{eqnarray} \label{as_4loop}
a_A(\mu_r) &=& \frac{1}{\beta_0^A L_A}
\left[ 1 - \frac{\beta_1^A \ln (L_A)}{{\beta_0^A}^2 L_A}
+ \left[ {\beta_1^A}^2 \left[ \ln^2 (L_A) - \ln (L_A) - 1 \right]
+ \beta_0^A \beta_2^A \right] \frac{1}{{\beta_0^A}^4 {L_A}^2}
\right. \nonumber \\
&& \left. - \left[ {\beta_1^A}^3 \left[ \ln^3 (L_A) - \frac{5}{2} \ln^2 (L_A) - 2 \ln (L_A) + \frac{1}{2} \right] + 3 \beta_0^A \beta_1^A \beta_2^A \ln (L_A) - \frac{1}{2} {\beta_0^A}^2 \beta_3^A \right] \frac{1}{{\beta_0^A}^6 {L_A}^3} \right] + {\cal O}\left(\frac{1}{L^5_A} \right),
\end{eqnarray}
\end{widetext}
where $L_{\rm A}=\ln \left({\mu_r^2}/{\Lambda_{\rm A}^2} \right)$ and $\Lambda_{\rm A}$ is the asymptotic scale under an arbitrary $\mathrm{A}$-scheme. The $\{\beta^{\mathrm{A}}_i\}$-functions under the $\mathrm{\overline{\rm MS}}$-scheme can be found in Refs.\cite{O.V. Tarasov, S.A.Larin, T.van, K.G.Chetyrkin, M.Czakon}. Under the Landau gauge, the $\{\beta^{\mathrm{A}}_i\}$-functions under the mMOM-scheme up to four-loop level are~\cite{L.von Smekal}
\begin{eqnarray} \label{batamom}
\beta^{\rm mMOM}_{0}&=& 11 -0.66667 n_f , \nonumber\\
\beta^{\rm mMOM}_{1}&=& 102 -12.66667 n_f , \nonumber\\
\beta^{\rm mMOM}_{2}&=& 3040.48229 -625.38667 n_f +19.3833 n_f^2 , \nonumber\\
\beta^{\rm mMOM}_{3}&=& 100541.0586 -24423.33055 n_f \nonumber\\
&& +1625.40224 n_f^2-27.49264 n_f^3,
\end{eqnarray}
where $n_f$ stands for the number of active flavors. Thus the $\alpha_s$-running is fixed if we know the asymptotic parameter $\Lambda_{\mathrm {mMOM}}$ well, whose value can be determined by using its relation to $\Lambda_{\overline{\rm MS}}$.

It is noted that both $\beta^{\rm mMOM}_0$ and $\beta^{\rm mMOM}_1$ are the same as those of $\mathrm{\overline{MS}}$-scheme, then, we can apply the extended renormalization group method suggested by Ref.\cite{extendedgroup} to evolve the $\mathrm{mMOM}$-running coupling $a_{\mathrm{mMOM}}(\mu_{r,\mathrm{mMOM}})$ at the scale $\mu_{r, \mathrm{mMOM}}$ ``adiabatically" into the $\overline{\mathrm{MS}}$-running coupling $a_{\mathrm{\overline{MS}}}(\mu_{r, \overline{\mathrm{MS}}})$ at the scale $\mu_{r,\overline{\mathrm{MS}}}$ not only in scale but also in scheme. Thus a more reasonable and accurate perturbative expansion of $a_{\mathrm{\overline{MS}}}(\mu_{r,\overline{\mathrm{MS}}})$ over $a_{\mathrm{mMOM}}(\mu_{r,\mathrm{mMOM}})$ can be achieved~\cite{Lu:1992nt}. As a special case, when taking the same arguments for the running coupling under different schemes, i.e. $\mu_{r,\overline{\mathrm{MS}}} =\mu_{r,\mathrm{mMOM}}\equiv\mu_r$, we get a perturbative series
\begin{eqnarray} \label{amomams3}
a_{\mathrm{\overline{MS}}}(\mu_r) &=&\sum^{\infty}\limits_{i=1} r_{i}\; a^{i}_{\mathrm{mMOM}}(\mu_r),
\end{eqnarray}
whose first three coefficients are
\begin{eqnarray}
r_1 &=& 1,  \label{relat1} \\
r_2 &=& -2{\beta_0}\ln\frac{\Lambda_{\mathrm {mMOM}}}{\Lambda_{\overline{\mathrm {MS}}}}, \label{relat2} \\
r_3 &=& \frac{\beta_2^{\overline{\mathrm {MS}}}}{\beta_0}-\frac{\beta_2^{\mathrm {mMOM}}}{\beta_0}-2{\beta_1}\ln\frac{\Lambda_{\mathrm {mMOM}}}{\Lambda_{\overline{\mathrm {MS}}}}+4{\beta_0^2}\ln^2\frac{\Lambda_{\mathrm {mMOM}}}{\Lambda_{\overline{\mathrm {MS}}}}. \label{relat3}
\end{eqnarray}

On the other hand, the coefficients $r_i$ have been directly calculated up to four-loop level~\cite{O.V.Tarasov, A.I.Davydychev, K.G.Chetyrkin3, L.von Smekal, J.A.Gracey2}
\begin{eqnarray}
r_{1}&\equiv& D_1= 1,  \label{amoms1} \\
r_{2}&\equiv& D_2= -14.0833 + 1.11111n_f, \label{amoms2} \\
r_{3}&\equiv& D_3= -78.7945 + 9.862n_f + 1.23457n_f^2, \label{amoms3} \\
r_{4}&\equiv& D_4= 862.512 -328.144 n_f + 19.256 n_f^2 + 1.37174n_f^3. \label{amoms4}
\end{eqnarray}

As a combination of Eqs.(\ref{relat1}, \ref{relat2}, \ref{relat3}) and Eq.(\ref{amoms1}, \ref{amoms2}, \ref{amoms3}), we obtain the wanted relations among the asymptotic scales $\Lambda_{\rm mMOM}$ and $\Lambda_{\mathrm{\overline{MS}}}$ up to three-loop level, i.e.
\begin{eqnarray}
&& \frac{\Lambda_{\rm mMOM}}{\Lambda_{\mathrm{\overline{MS}}}}\nonumber\\
&& = \exp\left[-\frac{D_{2}}{2\beta_0}\right] \label{lamda1} \\
&& =\exp\left[\frac{\beta_1- \sqrt{4D_3\beta^2_0 +\beta^2_1+4{\beta_0}({\beta_2^{\mathrm {mMOM}}} -\beta_2^{\overline{\mathrm {MS}}})}}{4\beta_0^2}\right] \label{lamda2} \\
&& = \cdots ,
\end{eqnarray}
where the symbol $\cdots$ stands for higher-order equations derived through a four-loop and even higher level comparisons. The first equation (\ref{lamda1}) is derived from a two-loop comparison, which equals to the one given by Refs.\cite{L.von Smekal, J.A.Gracey2} that is derived via a different approach. As a byproduct, those relations could be treated inversely as a consistency check of the complex high-order coefficients $D_i$ calculated in the literature. For example, the equivalence of Eq.(\ref{lamda1}) and Eq.(\ref{lamda2}) can be used as a cross-check of the correctness of $D_2$ and $D_3$.

\subsection{The PMC analysis of $H\to gg$ under the mMOM-scheme up to order $\alpha^5_s$}

Under the usual $\mathrm{\overline{MS}}$-scheme, the pQCD prediction for the total decay width $\Gamma(H\to gg)$ up to $n_\mathrm{th}$-loop level can be written as
\begin{equation}\label{fggms}
\Gamma(H\to gg) = \sum_{i=1}^n \mathcal{C}^{\mathrm{\overline{MS}}}_i(\mu_r)\; a_{\mathrm{\overline{MS}}}^{i+1}(\mu_r),
\end{equation}
where $\mu_r$ stands for an arbitrary choice of renormalization scale, and $\mathcal{C}^{\mathrm {\overline{MS}}}_i$ are $i_{\rm th}$-loop coefficients under the $\mathrm{\overline{MS}}$-scheme, whose values up to four-loop level can be found in Ref.\cite{P.A.Baikov}. With the help of Eqs.(\ref{amomams3}, \ref{amoms1}, \ref{amoms2}, \ref{amoms3}, \ref{amoms4}), it can be transformed into the mMOM-scheme as
\begin{eqnarray}
\Gamma(H\to gg) &=& \sum_{i=1}^n \mathcal{C}^{\rm mMOM}_i(\mu_r)\; a_{\rm mMOM}^{i+1}(\mu_r) \nonumber\\
&=& \frac{G_FM^3_H}{36\sqrt{2}\pi} \sum^{4}_{i=1} \left(\sum_{j=0}^{i-1} c_{i,j} n_f^j\right) a^{i+1}_{\mathrm{mMOM}}(\mu_r) \nonumber\\
&& +{\cal O}\left(a^6_{\mathrm{mMOM}}(\mu_r)\right),
\end{eqnarray}
where $G_F$ is the Fermi constant. For convenience, we put the mMOM-scheme $c_{i,j}$ coefficients with explicit scale dependence in the Appendix. As a further step, following the PMC $R_\delta$-scheme~\cite{PMC4, PMC5}, we can transform the $n_f$-series into the following $\{\beta^{\rm mMOM}_i\}$-power series,
\begin{widetext}
\begin{eqnarray} \label{fmominit}
\Gamma(H\to gg) &=&\frac{G_FM^3_H}{36\sqrt{2}\pi} \left[r_{1,0} a^2_{\mathrm{mMOM}}(\mu_r) +(r_{2,0}+2\beta_0^{\rm mMOM} r_{2,1}) a^3_{\mathrm{mMOM}}(\mu_r) +(r_{3,0}+2\beta_1^{\rm mMOM} r_{2,1}+3\beta_0^{\rm mMOM} r_{3,1} \right. \nonumber\\
&&\left. +3(\beta^{\rm mMOM}_0)^2 r_{3,2}) a^4_{\mathrm{mMOM}}(\mu_r) +(r_{4,0}+2\beta_2^{\rm mMOM} r_{2,1}+3\beta_1^{\rm mMOM} r_{3,1} + 7\beta_1^{\rm mMOM} \beta_0^{\rm mMOM} r_{3,2} \right. \nonumber\\
&&\left. +4\beta_0^{\rm mMOM} r_{4,1}+6(\beta^{\rm mMOM}_0)^2 r_{4,2}+4(\beta^{\rm mMOM}_0)^3 r_{4,3}) a^5_{\mathrm{mMOM}}(\mu_r) \right]+{\cal O} \left(a^6_{\mathrm{mMOM}}(\mu_r)\right).
\end{eqnarray}
\end{widetext}
The mMOM-scheme conformal coefficients $r_{i,0}$ with $i=(1,2,3,4)$ and the non-conformal coefficients $r_{i,j}$ with $1\leq j<i\leq4$ are also put in the Appendix.

The non-conformal $\{\beta^{\rm mMOM}_i\}$-terms determine the optimal behavior of the running coupling via RGE. After applying the PMC, i.e. by absorbing/resumming all $\{\beta^{\rm mMOM}_i\}$-terms into the running coupling, the pQCD series (\ref{fmominit}) can be finally simplified into the following scheme-independent conformal series:
\begin{eqnarray}\label{fmompmc}
\Gamma(H\rightarrow gg)&=& \frac{G_FM^3_H}{36\sqrt{2}\pi} \sum_{i=1}^{4} r_{i,0}\; a^{i+1}_{\mathrm{mMOM}}(Q_i) +\cdots.
\end{eqnarray}
Here $Q_i$ stands for the PMC scale at each perturbative order, whose values are determined by RGE and are usually different at different orders. Up to four-loop QCD corrections, the LO PMC scale $Q_1$, the NLO PMC scale $Q_2$ and the $\mathrm{N^2}$LO PMC scale $Q_3$ are
\begin{widetext}
\begin{eqnarray}
\ln\frac{Q_1^2}{\mu_r^2} &=& -\frac{r_{2,1}}{r_{1,0}}+ \frac{3(r_{2,1}^2-r_{1,0}r_{3,2})\beta^{\rm mMOM}_0}{2r_{1,0}^2} a_{\mathrm{mMOM}}(\mu_r)\nonumber\\
&& + \frac{(-5r_{2,1}^3+9r_{1,0}r_{2,1}r_{3,2}- 4r_{1,0}^2r_{4,3}) (\beta^{\rm mMOM}_0)^2 +4r_{1,0}(r_{2,1}^2 -r_{1,0}r_{3,2}) \beta^{\rm mMOM}_1} {2r_{1,0}^3} a_{\mathrm{mMOM}}^2(\mu_r)+{\cal O}(a_{\mathrm{mMOM}}^3), \label{pmcQ1} \\
\ln\frac{Q_2^2}{\mu_r^2} &=& -\frac{r_{3,1}}{r_{2,0}}+ \frac{2(r_{3,1}^2-r_{2,0}r_{4,2})\beta^{\rm mMOM}_0} {r_{2,0}^2}a_{\mathrm{mMOM}}(\mu_r)+{\cal O}(a_{\mathrm{mMOM}}^2), \label{pmcQ2} \\
\ln\frac{Q_3^2}{\mu_r^2} &=& -\frac{r_{4,1}}{r_{3,0}} + {\cal O}(a_{\mathrm{mMOM}}). \label{pmcQ3}
\end{eqnarray}
\end{widetext}
At the four-loop level, the N$^3$LO PMC scale $Q_4$ is undetermined due to unknown higher-order $\{\beta^{\rm mMOM}_i\}$-terms, and we set its value as the latest determined PMC scale $Q_3$. After applying the PMC, there are two kinds of residual scale dependence. The PMC scales are in perturbative series such that to eliminate all known non-conformal $\beta$-terms properly via RGE. The first one is from the PMC scales themselves, which are due to ``uncalculated" high-order $\beta$-terms. The second one is from the final perturbative term, in which we have no $\beta$-terms to set its PMC scale and we usually take it as the final PMC scale determined at one-order lower. Generally, those two residual scale dependence are highly suppressed, and the conventional renormalization scale uncertainty to the total and individual decay widths at each order can thus be greatly suppressed.

\section{Numerical results and discussions} \label{sect3}

\begin{table}[htb]
\centering
\begin{tabular}{ccccc}
\hline
~~&~1-loop~&~2-loop~&~3-loop~&~4-loop~\\
\hline
~$\Lambda_{n_f=5}^{\rm \overline{MS}}$~(GeV)~&~0.0904~&~0.233~&~0.214~&~0.214~\\
~$\Lambda_{n_f=5}^\mathrm{mMOM}$~(GeV)~&~0.158~&~0.406~ &~0.373~&~0.373~ \\
\hline
\end{tabular}
\caption{The asymptotic scales $\Lambda_{n_f=5}^{\rm \overline{MS}}$ and $\Lambda_{n_f=5}^\mathrm{mMOM}$ at different loop-levels under the $\overline{\rm MS}$-scheme and the mMOM-scheme. } \label{lam_scale}
\end{table}

To do the numerical calculation, we adopt $G_F=1.16638\times10^{-5}\; \mathrm{GeV}^{-2}$, the Higgs mass $M_H$=126 GeV, and the top-quark pole mass $m_t=173.3$ GeV~\cite{toppole}. The QCD asymptotic scales are determined by using the world average of the running coupling at the scale $M_Z$, $\alpha_s(M_Z)$=0.1185~\cite{PDG}. The asymptotic QCD scales under both the $\overline{\mathrm{MS}}$-scheme and the mMOM-scheme for $n_f=5$ at different loop-levels are presented in Table \ref{lam_scale}. The asymptotic scales with different flavors can be determined via the usual matching, cf. Ref.\cite{alphasrunning}. Moreover, to be self-consistent, we shall adopt the $n_\mathrm{th}$-loop $\alpha_s$-running to predict the decay width $\Gamma(H\to gg)$ up to $n_\mathrm{th}$-loop QCD corrections.

\subsection{Total decay width $\Gamma(H\to gg)$ at $\alpha_s^5$-order before and after the PMC scale-setting}

\begin{table}[htb]
\centering
\begin{tabular}{cccccc}
\hline
~ & ~LO~ & ~NLO~ & ~$\mathrm{N^2}$LO~ & ~$\mathrm{N^3}$LO~ & ~Total~ \\
\hline
~~$\Gamma_i|_{M_H/2}$~~ &284.0&104.5&-28.2&-17.5&~~342.8 \\
$\Gamma_i|_{M_H}$&224.6&120.4&15.0&-9.1&~~350.9 \\
$\Gamma_i|_{2M_H}$&184.4&124.3&41.9&5.5&~~356.1 \\
$\Gamma_i|_{4M_H}$&155.4&123.2&59.5&20.1&~~358.2 \\
\hline
\end{tabular}
\caption{Total and individual decay widths (in unit: keV) of $H\to gg$ at $\alpha_s^5$-order under conventional scale-setting and the mMOM-scheme. $\Gamma_i$ stands for the individual decay width at each order with $i={\rm LO}$, ${\rm NLO}$, ${\rm N^2LO}$, and ${\rm N^3LO}$, respectively; $\Gamma_{\rm Total}=\sum_{i}\Gamma_{i}$ stands for total decay width. $\mu_r=M_H/2$, $M_H$, $2M_H$, and $4M_H$, respectively. } \label{widthCON}
\end{table}

We present the total and individual decay widths of $H\to gg$ at $\alpha_s^5$-order under conventional scale-setting and the mMOM-scheme in Table \ref{widthCON}, in which the results are given by varying $\mu_{r} \in[M_H/2, 4M_H]$. $\Gamma_{\rm Total}=\sum_{i}\Gamma_{i}$ stands for total decay width, where $\Gamma_i$ stands for the individual decay width at each order with $i={\rm LO}$, ${\rm NLO}$, ${\rm N^2LO}$, and ${\rm N^3LO}$, respectively. It is found that under conventional scale-setting, the total decay width $\Gamma_{\rm Total}$ increases with the increment of $\mu_r$ for $\mu_r \precsim 450$ GeV, and then it slightly decreases with the increment of $\mu_r$, which changes down to $355$ keV for $\mu_r=1 \mathrm{TeV}$. Moreover, for the case of $\mu_r=M_H$, Table \ref{widthCON} shows that the $K$ factor under the mMOM-scheme follows the trends
\begin{equation}
K_{\rm mMOM}|_{\rm Conv.} \sim 1+0.54+0.07-0.04 .
\end{equation}
In comparison to Eq.(\ref{msbarK}), it indicates that a more convergent pQCD series than that of $\overline{\rm MS}$-scheme can be achieved by using the mMOM-scheme even before applying the PMC scale-setting. After applying the PMC, the pQCD series shall be further improved as
\begin{eqnarray}
K_{\rm mMOM}|_{\rm PMC} \sim 1+0.30 -0.10-0.04 .
\end{eqnarray}

Under conventional scale-setting, a single renormalization scale is ``guessed", which, at high orders, may give a reasonable prediction for a global observable such as the total decay width or total cross section that is close to the experimental result; however, the corresponding predictions for more detailed observables such as the correlations could be inaccurate or even wrong. As an example: The two-loop prediction for the total cross-section of the top-pair production agrees with the CDF and D0 measurements by simply setting the scale as the top-quark mass; however one then finds a large discrepancy with the top-pair forward-backward asymmetry measured at the Tevatron. In contrast, the PMC prediction predicts both the total cross-section and the forward-backward asymmetry correctly~\cite{pmc1, pmc3, pmc4, Wang:2015lna}, which is also confirmed by the PMC analysis of top-pair production at the LHC~\cite{pmcLHC}.

Table \ref{widthCON} shows that by varying $\mu_{r} \in[M_H/2, 4M_H]$, the total decay width under conventional scale-setting shall be changed by about $\pm 2\%$ from its central value at $\mu_{r}=M_H$. This indicates that the scale dependence for the total decay width are small at the present $\alpha_s^5$-order. By analyzing the pQCD series in detail, it is found that the scale errors are rather large for each perturbative term $\Gamma_i$, thus such a small scale error ($\pm 2\%$) for the total decay width is due to cancelations among different orders. For definiteness, we define a ratio to show explicitly how the individual decay width $\Gamma_i$ at each order changes with different choices of scale, i.e.,
\begin{equation}
\kappa_{i}=\frac{\Gamma_{i}|_{\mu_r}-\Gamma_{i}|_{\mu_{r}\equiv M_H}} {\Gamma_{i}|_{\mu_{r}\equiv M_H}}\times 100\% , \label{kappadef}
\end{equation}
where $\Gamma_i$ stands for the individual decay width at each order and $i={\rm LO}$, ${\rm NLO}$, ${\rm N^2LO}$, and ${\rm N^3LO}$, respectively. Under conventional scale-setting, we get
\begin{eqnarray}
&& \kappa_{\rm LO}=26\%, \; \kappa_{\rm NLO}=-13\%, \nonumber\\
&& \kappa_{\rm N^2LO}=-288\%, \; \kappa_{\rm N^3LO}=92\% \label{kappaCON1}
\end{eqnarray}
for $\mu_r=M_H/2$; we get
\begin{eqnarray}
&& \kappa_{\rm LO}=-18\%, \; \kappa_{\rm NLO}=3\%, \nonumber\\
&& \kappa_{\rm N^2LO}=179\%, \; \kappa_{\rm N^3LO}=-160\% \label{kappaCON2}
\end{eqnarray}
for $\mu_r=2M_H$. Those large $\kappa_i$ values indicate that under conventional scale-setting, one cannot decide what is the exact decay width for each perturbative order.

\begin{table}[htb]
\centering
\begin{tabular}{cccccc}
\hline
~ & ~LO~ & ~NLO~ & ~$\mathrm{N^2}$LO~ & ~$\mathrm{N^3}$LO~ & ~Total~ \\
\hline
~~$\Gamma_i|_{M_H/2}$~~ & 298.0 &87.9&-30.0&-12.5&~~343.4\\
$\Gamma_i|_{M_H}$ & 298.2&90.2 &-30.0&-12.5&~~345.9\\
$\Gamma_i|_{2M_H}$ & 298.5&91.9 &-30.0&-12.5&~~347.9\\
$\Gamma_i|_{4M_H}$ & 298.8&93.3 &-30.0&-12.5&~~349.6\\
\hline
\end{tabular}
\caption{The PMC predictions for the total and individual decay widths (in unit: keV) of $H\to gg$ at $\alpha_s^5$-order under the mMOM-scheme. $\Gamma_i$ stands for the individual decay width at each order with $i={\rm LO}$, ${\rm NLO}$, ${\rm N^2LO}$, and ${\rm N^3LO}$, respectively; $\Gamma_{\rm Total}=\sum_{i}\Gamma_{i}$ stands for total decay width. $\mu_r=M_H/2$, $M_H$, $2M_H$, and $4M_H$, respectively. } \label{widthPMC}
\end{table}

As a comparison, we present the PMC prediction for the total and individual decay widths under the mMOM-scheme in Table \ref{widthPMC}, in which the results are given by taking $\mu_{r}=$ $M_H/2$, $M_H$, $2M_H$ and $4M_H$, respectively. Table \ref{widthPMC} shows by varying $\mu_{r} \in[M_H/2, 4M_H]$, the total decay width under PMC scale-setting shall only be changed by about $\pm1\%$. Moreover, the renormalization scale errors for the individual decay widths $\Gamma_i$ have also been greatly suppressed by PMC scale-setting, e.g. except for the NLO ratio $|\kappa_{\rm NLO}| < 4\%$, all other $\kappa_i$ are less than $0.2\%$ by varying $\mu_r$ within the region of $\in[M_H/2, 4M_H]$. This can be explained by the fact that after applying the PMC, we can fix the $\alpha_s$-running behavior, or equivalently the renormalization scale, at each perturbative order via the using of RGE.

\begin{figure}[tb]
\centering
\includegraphics[width=0.50\textwidth]{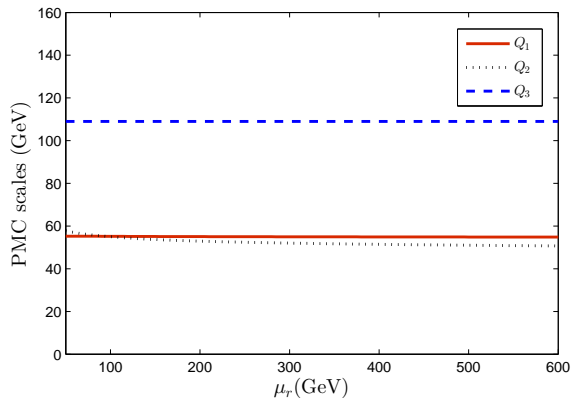}
\caption{The LO, NLO and $\mathrm{N^2}$LO PMC scales $Q_1$, $Q_2$ and $Q_3$ versus the initial choice of scale $\mu_r$ for $H\to gg$ at $\alpha_s^5$-order under the mMOM-scheme, which are shown by solid, dotted and dashed lines, respectively.} \label{scalesq1q2q3}
\end{figure}

More specifically, for the present process at the $\alpha_s^5$-order, we have three PMC scales $Q_{1,2,3}$, which are determined by Eqs.(\ref{pmcQ1}, \ref{pmcQ2}, \ref{pmcQ3}). They are functions of the initial scale $\mu_r$. Generally, the initial scale $\mu_r$ can be chosen arbitrarily, which only needs to be within the perturbative region to ensure the reliability of a pQCD calculation. The PMC adopts the $\{\beta^{\rm mMOM}_i\}$-terms, determined by RGE, and fixes the optimal scales of the process. We present the PMC scales versus the initial scale $\mu_r$ in Fig.(\ref{scalesq1q2q3}). Those optimal PMC scales are smaller than the conventional choice of scale $(M_H)$ to a certain degree. Fig.(\ref{scalesq1q2q3}) shows that the PMC scales $Q_{1,2,3}$ are highly independent on the choice of $\mu_r$, and their values are almost fixed to be
\begin{eqnarray}
Q_1\simeq 55{\rm GeV},\; Q_2\simeq 54{\rm GeV},\; Q_3\simeq 109{\rm GeV}.
\end{eqnarray}
This explains why the individual decay widths at each order and the total decay width are highly independent to the initial choice of scale. Thus the conventional scale ambiguity is cured.

The $\{\beta^{\rm mMOM}_i\}$-terms are governed by RGE, which are generally different at different orders, thus the PMC scales $Q_i$ at different orders are different from each other~\cite{PMC2}. The PMC scales themselves are in perturbative series~\cite{PMC4, PMC5}, i.e., for $H\to gg$ at order $\alpha_s^5$, the LO PMC scale $Q_1$ is at the accuracy of next-to-next-to-leading logarithmic order (NNLLO); the NLO PMC scale $Q_2$ is at the accuracy of NLLO; and the N$^2$LO PMC scale $Q_3$ is at the accuracy of LLO. This way, the PMC scales can be improved when more-and-more QCD loop corrections are considered, and the unknown higher-order $\{\beta^{\rm mMOM}_i\}$-terms shall lead to residual scale dependence. In principal, as shown by Refs.\cite{pmc1, pmc3, pmc4, Wang:2015lna, pmcLHC, app1, app2, Shen:2015cta}, due to both the $\alpha_s$-power suppression and the exponential-suppression, such residual scale dependence is negligible. For the present case, we observe that there is a somewhat larger residual scale uncertainty in comparison to previous PMC applications, i.e. a sizable $\sim1\%$ residual scale dependence has been found to total decay width. This indicates that there are large contributions from the unknown $\{\beta^{\rm mMOM}_i\}$-terms. More explicitly, by using the formulas (\ref{pmcQ1}, \ref{pmcQ2}, \ref{pmcQ3}), we obtain
\begin{eqnarray}\label{pmcselfs}
\ln Q_1^2/\mu_r^2|_{\mu_r=M_H} &\simeq& -1.83 + 0.51 + 0.19 + {\cal O}(\alpha_s^3), \\
\ln Q_2^2/\mu_r^2|_{\mu_r=M_H} &\simeq& -2.42 + 2.29 + {\cal O}(\alpha_s^2), \label{pmcselfs22} \\
\ln Q_3^2/\mu_r^2|_{\mu_r=M_H} &\simeq& -0.29 + {\cal O}(\alpha_s).
\end{eqnarray}
The PMC scale $Q_1$ shows a good pQCD convergence as the usual PMC applications. However, the pQCD convergence of $Q_2$ is questionable, the magnitude of its NLO term is at the same order of the LO term, which explains the larger residual scale dependence shown by Fig.(\ref{scalesq1q2q3}). It also explains why there is somewhat larger residual scale dependence to $\Gamma_{\rm NLO}$, as shown by Table~\ref{widthPMC}. Thus to further improve our PMC predictions, we need to finish even higher-order corrections, such as a five-loop calculation, so as to achieve an accurate NLO PMC scale $Q_2$ and a precise prediction of the NLO decay width $\Gamma_{\rm NLO}$ with less residual scale dependence.

\subsection{Total decay width up to different $\alpha_s$-orders}

In above subsection, we have shown the properties of total decay width at order $\alpha^5_s$ under the mMOM-scheme. As a step forward, it is helpful to learn how the total decay width $\Gamma(H\to gg)$ behaves before and after the PMC scale-setting when more-and-more loop terms are included. In present subsection, we make a detailed discussion on the properties of total decay width from one-loop level to four-loop level.

\begin{figure}[tb]
\centering
\includegraphics[width=0.50\textwidth]{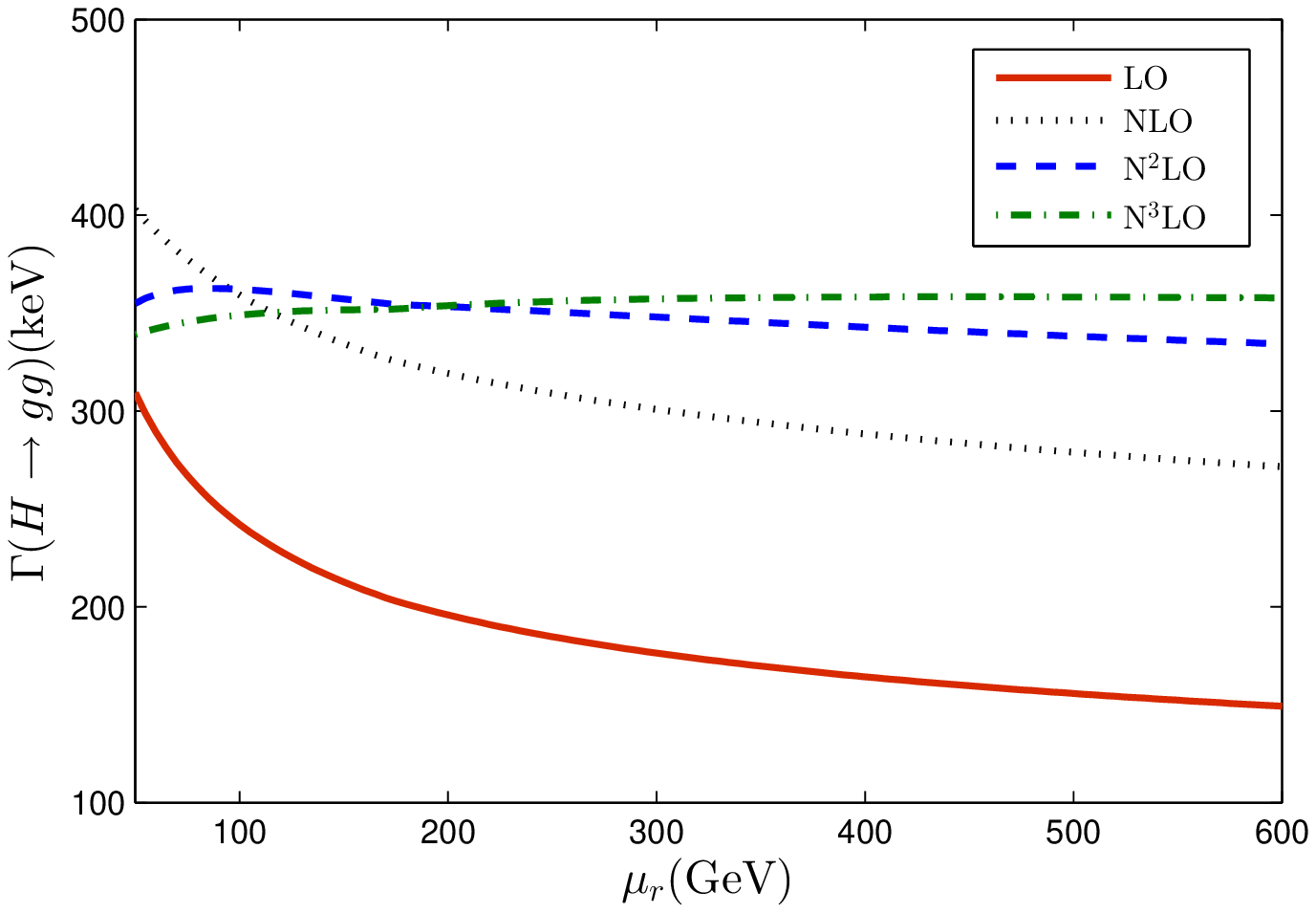}
\includegraphics[width=0.50\textwidth]{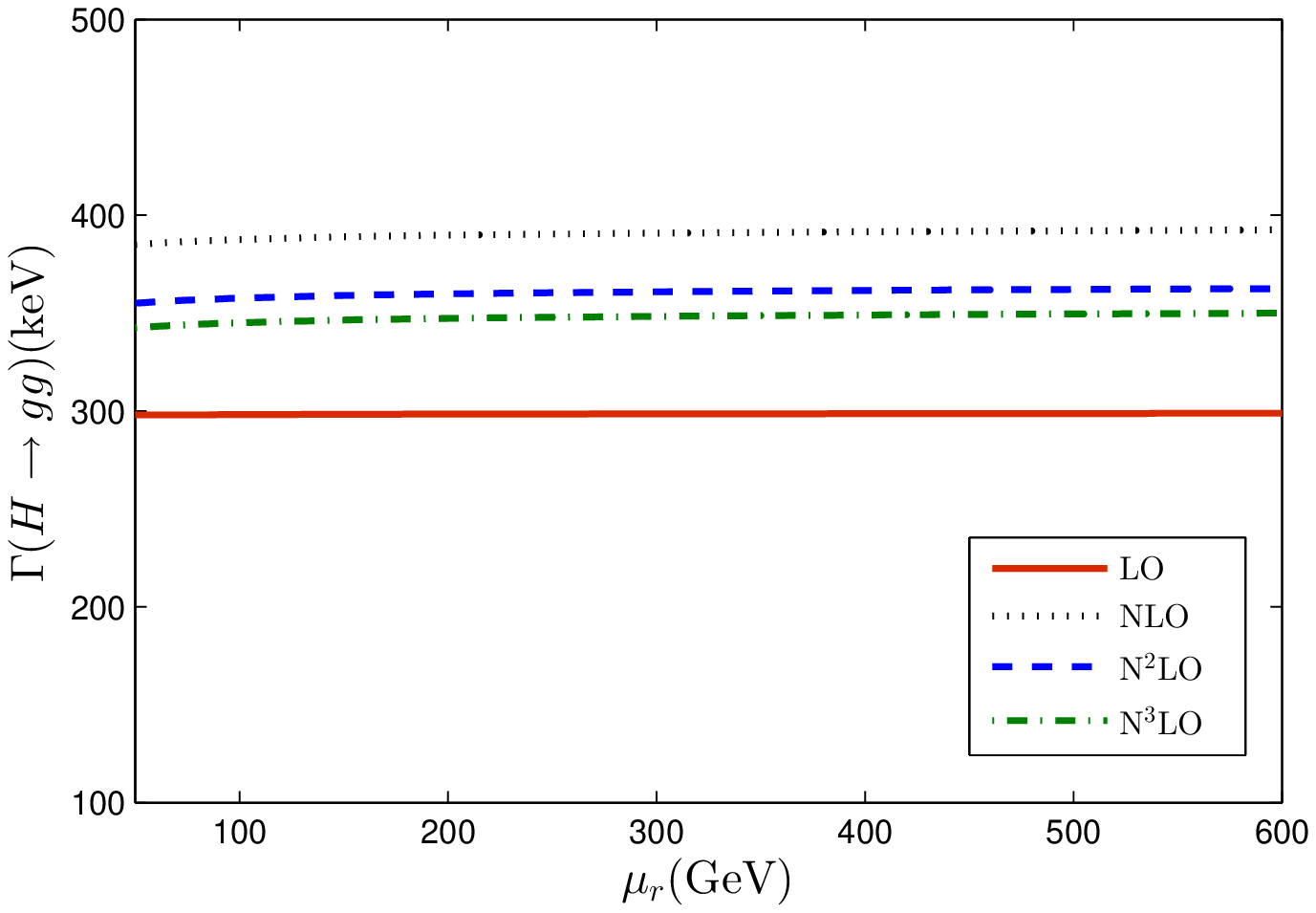}
\caption{Total decay width $\Gamma(H\to gg)$ versus the scale $\mu_r$ up to four-loop level under the mMOM-scheme. The solid, the dotted, the dashed and the dash-dot lines are for LO, NLO, $\mathrm{N^2}$LO and $\mathrm{N^3}$LO total decay width $\Gamma(H\to gg)$, respectively. The upper diagram is for conventional scale-setting, and the lower one is for PMC scale-setting.}
\label{width1}
\end{figure}

We present the total decay width $\Gamma(H\to gg)$ under the mMOM-scheme versus the initial scale $\mu_r$ up to four-loop level before and after the PMC scale-setting in Fig.(\ref{width1}). Fig.(\ref{width1}) shows that under conventional scale-setting, the LO and NLO total decay width $\Gamma(H\to gg)$ depend heavily on $\mu_r$, which then becomes weaker-and-weaker when more-and-more loop corrections are taken into consideration; and at the four-loop level, the scale uncertainty of $\Gamma(H\to gg)$ is about $\pm 2\%$ within scale region $\mu_r\in[M_H/2,4M_H]$. This agrees with the conventional wisdom that by finishing a higher-and-higher order calculation, one can get a desirable scale-invariant estimate. However, under conventional scale-setting, by using a single ``guessed" scale, the scale ambiguities and scheme-dependence persist at any fixed order: if one uses conventional scale-setting for an $\alpha_s^{n}$-order pQCD prediction, the large scale ambiguity exists for any known perturbative terms, as has been shown by the last subsection. On the other hand, it is found that after applying the PMC scale-setting, the total decay width $\Gamma(H\to gg)$ with QCD corrections up to LO, NLO, $\rm N^{2}LO$ and $\rm N^{3}LO$, accordingly, are almost flat versus the scale $\mu_{r}$. Fig.(\ref{width1}) also shows that after the PMC scale-setting, the value of $\Gamma(H\to gg)$ shows a faster steady behavior than the conventional predictions by including higher-and-higher order corrections, e.g. it quickly approaches its steady value with more-and-more loop corrections included.

As shown by Tables \ref{widthCON} and \ref{widthPMC}, if setting $\mu_r=M_H/2$ for conventional scale-setting, we get almost the same total decay width as the PMC one. This indicates that for conventional scale-setting, the best choice for $H\to gg$ should be $\sim M_H/2$ other than the usually suggested $M_H$. Similar choice has been tried for analyzing the gluon-fusion channel $gg\to H$~\cite{ggtoh1, ggtoh2, ggtoh3}, which has the same topology as the decay $H\to gg$. In Refs.\cite{ggtoh1, ggtoh2, ggtoh3}, this choice of scale is ``guessed", and our present derivation provides a reason for this choice~\footnote{A detailed PMC analysis to the important Higgs hadro-production channel, $gg\to H$, is in preparation.}.

It is helpful to estimate the magnitude of ``unknown" higher-order pQCD prediction. We adopt the way suggested by Ref.\cite{Wu:2014iba} for such an estimation, i.e. for a $n_{\rm th}$-loop pQCD prediction
\begin{eqnarray}
\Gamma^{\rm tot}_n &=& \sum_{i=1}^n \mathcal{C}^{\rm mMOM}_i(\mu_r) a_{\rm mMOM}^{i+1}(Q_i[\mu_r]),
\end{eqnarray}
whose perturbative uncertainty is
\begin{equation}
\Delta\Gamma^{\rm tot}_n =\pm|{\cal C}^{\rm mMOM}_{n}(\mu_r) a_{\rm mMOM}^{n+1}(Q_n[\mu_r])|_{\rm MAX},
\end{equation}
which is calculated by varying $\mu_{r}$ within the region of $[M_H/2, 4M_H]$, and the symbol ``MAX" stands for the maximum value of $|{\cal C}^{\rm mMOM}_i \; a_{\rm mMOM}^{i+1}|$ within this scale region. Under conventional scale-setting, $Q_i[\mu_r]\equiv \mu_r$; under PMC scale-setting, $Q_i[\mu_r]$ are PMC scales and $\mathcal{C}^{\rm mMOM}_i(\mu_r)\equiv {G_FM^3_H}/{(36\sqrt{2}\pi)} r_{i,0}$ are conformal coefficients. This way of estimating ``unknown" higher-order pQCD prediction is natural for PMC, since after the PMC scale-setting, the pQCD convergence is ensured and the only uncertainty is from the last term due to unfixed PMC scale at this particular order.

\begin{figure}[tb]
\centering
\includegraphics[width=0.50\textwidth]{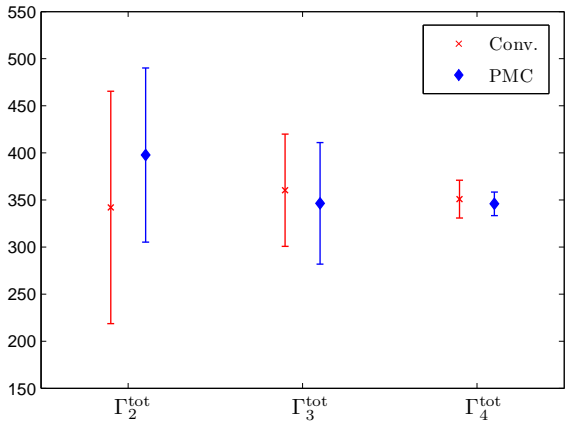}
\caption{Results of total decay width $\Gamma(H\to gg)=\Gamma^{\rm tot}_n$ (in unit: keV) at different pQCD orders with their errors $\Delta\Gamma^{\rm tot}_n$, $n=(2, 3, 4)$, respectively. The crosses and the diamonds are for the ones under conventional and PMC scale-settings, respectively. The central values are for $\mu_r=M_H$. }
\label{errors}
\end{figure}

The fixed-order prediction together with its errors on the total decay width $\Gamma(H\to gg)$ at different pQCD orders before and after the PMC scale-setting are displayed in Fig.(\ref{errors}), in which the ``Conv." stands for the result under conventional scale-setting. In both cases, the predicted error bars from ``unknown" higher-order terms show a better agreement with their steady points when more loop-terms are ``known".

To compare with the case of conventional scale-setting, after the PMC scale-setting, the predicted ``unknown" higher-order contributions are usually smaller than the conventional ones~\cite{Wu:2014iba}. For the present case, there is one exception that the predicted $\Delta\Gamma^{\rm tot}_3$ from a three-loop prediction $\Gamma^{\rm tot}_3$ is still large. Numerically it is caused by a smaller NLO PMC scale $Q_2$ $(<M_H)$ and a larger N$^2$LO conformal coefficient $[{G_FM^3_H}/{(36\sqrt{2}\pi)} r_{3,0}]$, which is about two times larger than the N$^2$LO perturbative coefficient $C^{\rm mMOM}_3$ \footnote{It is noted that a smaller coefficient $C^{\rm mMOM}_3$ in comparison to the conformal one is due to the accidentally large cancelation among the conformal and non-conformal terms.}. This exception does not indicate the breaking of the general features of the PMC scale-setting, which is consistent with the previous observation that the NLO PMC scale $Q^2$ is of large residual uncertainty. In fact, for the present decay channel, the predicted error bars under conventional scale-setting cannot be treated too seriously: the large $\kappa_i$ values, as shown by Eqs.(\ref{kappaCON1}, \ref{kappaCON2}), indicate that one cannot decide the exact values for all the individual $\Gamma_i$ under conventional scale-setting.

\section{Summary}
\label{sect4}

Following the renormalization group invariance, the physical observables must be independent on the choice of renormalization scheme. The PMC provides a solid and unambiguous procedure to set the renormalization scale for any QCD process at any finite order. One of its key point is to correctly deal with the $\{\beta_i\}$-terms. The physical mMOM-scheme is helpful for setting the $\{\beta_i\}$-terms of the process unambiguously, especially for the processes involving three-gluon or four-gluon vertex at lower orders. In the paper, we have made a detailed PMC analysis on the Higgs-boson decay $H\to gg$ within the mMOM-scheme up to four-loop QCD corrections.

It is the first time to apply the PMC to deal with N$^2$LO and higher order mMOM pQCD predictions. As indicated by Table~\ref{widthCON}, the pQCD convergence under the mMOM-scheme is better than that of the $\overline{\rm MS}$-scheme even before the PMC scale-setting. After applying the PMC, the pQCD convergence can be further improved.

Table \ref{widthCON} shows that the scale dependence for total decay width are small at the present $\alpha_s^5$-order even under conventional scale-setting, i.e. by varying $\mu_{r} \in[M_H/2, 4M_H]$, the total decay width shall only be changed by $\sim\pm 2\%$. Under the conventional scale-setting, by taking $M_H=126$ GeV, we obtain
\begin{equation}
\Gamma(H\to gg)|_{\rm mMOM, Conv.} =\sum_{i}\Gamma_i = 350.9^{+7.3}_{-8.1} \; {\rm keV},
\end{equation}
where the central value is for $\mu_{r}=M_H$ and the error is for $\mu_r\in[M_H/2,4M_H]$. As we have pointed out in the body of the text, the scale errors for each loop terms $\Gamma_i$ are quite large, and a small net scale error for the four-loop total decay width is due to the large cancelations among different orders. More explicitly, the scale errors for the decay widths $\Gamma_i$ can be reexpressed by the parameter $\kappa_i$ (defined in Eq.(\ref{kappadef})), which are $\kappa_{\rm LO}\in[26\%,-31\%]$, $\kappa_{\rm NLO}\in[-13\%,2\%]$, $\kappa_{\rm N^2LO}\in[-288\%,297\%]$ and $\kappa_{\rm N^3LO}\in[92\%,-321\%]$ for $\mu_r\in[M_H/2,4M_H]$.

After applying the PMC scale-setting, the scale errors for either the total decay width $\Gamma(H\to gg)$ or its individual contributions $\Gamma_i$ are largely suppressed. By taking $M_H=126$ GeV, we get
\begin{equation}
\Gamma(H\to gg)|_{\rm mMOM, PMC} = 345.9^{+3.7}_{-2.5} \; {\rm keV}, \label{PMCHgg}
\end{equation}
where the central value is for $\mu_{r}=M_H$ and the error is caused by varying $\mu_r\in[M_H/2,4M_H]$. The central decay width is lowered by about $1\%$ in comparison to conventional one. The scale error for the decay width $\Gamma_i$ of each loop is greatly suppressed, e.g. except for the NLO ratio $|\kappa_{\rm NLO}| \sim 4\%$, all other ratios $\kappa_i$ are less than $0.2\%$ by varying $\mu_r\in[M_H/2, 4M_H]$. It should be pointed out that the somewhat larger residual scale error as shown by Eq.(\ref{PMCHgg}) in comparison to previous PMC examples are due to the uncalculated high-order $\beta$-terms. As shown by Table~\ref{widthPMC}, such large error is from the residual scale dependence of the NLO decay width $\Gamma_{\rm NLO}$, whose PMC scale $Q_2$ has a poor pQCD convergence as shown by Eq.(\ref{pmcselfs22}). Thus to further improve the accuracy of the PMC predictions, one needs to finish even high-order corrections such as a five-loop calculation to the present channel so as to achieve an accurate $Q_2$ and a precise prediction of $\Gamma_{\rm NLO}$ with less residual scale dependence.

Finally, if setting the Higgs mass as the recently measured one by the ATLAS and CMS collaborations, the PMC prediction of the total decay width is
\begin{eqnarray}
\Gamma(H\rightarrow gg)|_{\rm mMOM, PMC} &=& 339.3\pm 1.7^{+3.7}_{-2.4} \; {\rm keV},
\end{eqnarray}
where the first error is caused by taking the Higgs mass $M_H=125.09\pm 0.21\pm 0.11$ GeV~\cite{higgsmass}, and the second error is caused by varying $\mu_r\in[M_H/2,4M_H]$. To compare with Eq.(\ref{PMCHgg}), it shows that a change of Higgs mass by 1 GeV, the total decay width shall be changed by about 6 keV. Thus the eliminating of the renormalization scale error shall inversely help us to get an accurate prediction on Higgs mass, and etc.. \\

\noindent{\bf Acknowledgement}: The authors would like to thank Lorenz von Smekal, Kim Maltman, Andre Sternbeck, and Xu-Chang Zheng for helpful discussions. This work is supported in part by Fundamental Research Funds for the Central Universities under Grant No.CDJZR10100023 and No.CDJZR305513, and by the National Natural Science Foundation of China under Grant No.11275280, No.11547010 and No.11547305.

\appendix

\section*{Appendix: The coefficients $c_{i,j}$ and $r_{i,j}$ under the mMOM-scheme}

As mentioned in the body of the text, the $c_{i,j}$ coefficients under the mMOM-schemes can be obtained from those of $\mathrm{\overline{MS}}$-ones with the help of Eq.(\ref{amomams3}). The mMOM $c_{i,j}$ coefficients under Landau gauge with full renormalization scale dependence are
\begin{eqnarray}
c_{1,0} &=& 16, \\
c_{2,0} &=& 1069.33 - 352\ln\frac{M_H^2}{\mu_r^2},\\
c_{2,1} &=& -39.1111 + 21.3333\ln\frac{M_H^2}{\mu_r^2},\\
c_{3,0} &=& 31202.1 + 608\ln\frac{M_H^2}{m_t^2} - 38552\ln\frac{M_H^2}{\mu_r^2}\nonumber\\
&& + 5808\ln^2\frac{M_H^2}{\mu_r^2}, \\
c_{3,1} &=& -4043.54 + 170.667\ln\frac{M_H^2}{m_t^2} + 3834.67 \ln\frac{M_H^2}{\mu_r^2}\nonumber\\
&& - 704 \ln^2\frac{M_H^2}{\mu_r^2},\\
c_{3,2} &=& 41.2236 - 78.2222\ln\frac{M_H^2}{\mu_r^2} + 21.3333\ln^2\frac{M_H^2}{\mu_r^2},\\
c_{4,0} &=& -88214.8 + 6688\ln^2\frac{M_H^2}{m_t^2} + \ln\frac{M_H^2}{m_t^2}(34008.9 \nonumber\\
&& - 26752\ln\frac{M_H^2}{\mu_r^2}) - 1.7974\times10^6\ln\frac{M_H^2}{\mu_r^2} \nonumber\\
&& + 902000\ln^2\frac{M_H^2}{\mu_r^2} - 85184\ln^3\frac{M_H^2}{\mu_r^2}, \\
c_{4,1} &=& -128899 + 1472\ln^2\frac{M_H^2}{m_t^2} + \ln\frac{M_H^2}{m_t^2}(8037.93 \nonumber\\
&& - 5888\ln\frac{M_H^2}{\mu_r^2}) + 333737 \ln\frac{M_H^2}{\mu_r^2} \nonumber\\
&& - 145717\ln^2\frac{M_H^2}{\mu_r^2} + 15488\ln^3\frac{M_H^2}{\mu_r^2},\\
c_{4,2}&=& 3663.24 - 113.778 \ln^2\frac{M_H^2}{m_t^2} + \ln\frac{M_H^2}{m_t^2} (53.3333 \nonumber\\
&& + 455.111 \ln\frac{M_H^2}{\mu_r^2}) - 14703.1\ln\frac{M_H^2}{\mu_r^2} \nonumber\\
&& + 7239.11\ln^2\frac{M_H^2}{\mu_r^2} - 938.667 \ln^3\frac{M_H^2}{\mu_r^2},\\
c_{4,3}&=&9.97767 + 109.929\ln\frac{M_H^2}{\mu_r^2} - 104.296\ln^2\frac{M_H^2}{\mu_r^2} \nonumber\\
&& + 18.963\ln^3\frac{M_H^2}{\mu_r^2}.
\end{eqnarray}
There are two types of logarithmic terms $\ln{M_H^2}/{\mu_r^2}$ and $\ln{m_t^2}/{\mu_r^2}$ in those expressions, and we have used $\ln{M_H^2}/{m_t^2}$ to replace $\left(\ln{M_H^2} / {\mu_r^2} -\ln{m_t^2}/{\mu_r^2}\right)$ for brevity. \\

The expansion of the mMOM-scheme coefficients $r_{i,j}$ up to four-loop level under the Landau gauge are
\begin{eqnarray}
r_{1,0} &=& c_{1,0},\\
r_{2,1} &=& -\frac{3c_{2,1}}{2n},\\
r_{2,0} &=& c_{2,0}+\frac{33c_{2,1}}{2},\\
r_{3,2} &=& \frac{9c_{3,2}}{2n(1 + n)},\\
r_{3,1} &=& -\frac{6(c_{3,1}+33c_{3,2})-114c_{2,1}}{4(1+n)},\\
r_{3,0} &=& \frac{1}{8}(8c_{3,0}+66(2c_{3,1}+33c_{3,2}) -1284c_{2,1}),\\
r_{4,3} &=& -\frac{81 c_{4,3}}{4n(2+3n+n^2)},\\
r_{4,2} &=& \frac{1}{8(1+n)^2(2+n)}(989c_{2,1}(1+n) + 18((2c_{4,2} \nonumber\\
&& +99c_{4,3})(1+n)-38c_{3,2}(3+2n)) \nonumber\\
&& +48c_{2,1}(1+n)\zeta_3),\\
r_{4,1} &=& \frac{1}{32(1+n)(2+n)}(-14554c_{2,1}(1+n) \nonumber\\
&& -6(2(4c_{4,1}+33(4c_{4,2} + 99c_{4,3}))(1+n) \nonumber\\
&& -152c_{3,1}(1+n) + c_{3,2}(-8868 - 7584n)) \nonumber\\
&& -3162 c_{2,1}(1+n)\zeta_3),\\
r_{4,0} &=& \frac{1}{64}(3044c_{2,1} + 8(8c_{4,0} + 132c_{4,1} + 2178c_{4,2}\nonumber\\
&& + 35937c_{4,3} -1284c_{3,1} - 42372c_{3,2})),
\end{eqnarray}
where $\zeta_3$ is Riemann zeta-function, and $n$ stands for the LO $\alpha_s$-power, which equals to $2$ for the decay $H\to gg$.

\end{document}